%% file: main.tex
\documentclass[11pt]{article}
\pdfoutput=1 
\usepackage{palatino,microtype}
\usepackage[margin=15pt,font=small,labelfont={bf}]{caption}
\usepackage[margin=1in]{geometry}
\usepackage[english]{babel}
\usepackage[utf8]{inputenc}
\usepackage[compact]{titlesec}
\usepackage{cmap}
\usepackage[T1]{fontenc}
\usepackage{bm}
\usepackage[colorinlistoftodos]{todonotes}
\usepackage{booktabs}
\usepackage{mathtools}
\usepackage{amsmath}
\usepackage{amssymb}
\usepackage{amsthm}
\usepackage{float}
\usepackage{graphics}
\usepackage[numbers]{natbib}
\usepackage{hyperref}

\usepackage{colortbl}
\colorlet{myred}{red!25}
\colorlet{myblue}{blue!25}
\colorlet{mygreen}{green!25}

\newcommand{\BibTeX}{\rm B\kern-.05em{\sc i\kern-.025em b}\kern-.08em\TeX}

\usepackage[labelfont={normalfont,bf},textfont=it]{caption}
\usepackage{subcaption}

\usepackage{nicefrac}
\usepackage{pifont}

\usepackage{enumitem}
\usepackage{apxproof}
\usepackage{array,multirow,graphicx,bigdelim}
\input{insbox}
\usepackage{tikz}
\usepackage{tikz-dimline}
\usepackage{tikz-cd}
\usetikzlibrary{fit,calc,math,shapes,shapes.multipart,decorations.text,arrows,decorations.markings,decorations.pathmorphing,shapes.geometric,positioning,decorations.pathreplacing}

\tikzset{snake it/.style={decorate, decoration=snake}}

\tikzset{
    leading_agent/.style={circle, draw={rgb, 255:red, 93; green, 166; blue, 13}, 
                          fill={rgb, 255:red, 232; green, 254; blue, 212}, 
                          line width=0.75pt, minimum size=6mm, inner sep=0pt}, 
    agent/.style={circle, draw=black, fill=lightgray!30, line width=0.75pt, minimum size=6mm, inner sep=0pt}, 
    etc/.style={draw=none, minimum size=6mm, inner sep=0pt},
    envy/.style={-{Triangle}, line width=1pt}, 
    champ/.style={-{Triangle}, bend left=20, color={rgb, 255:red, 208; green, 2; blue, 27}, line width=1pt}, 
    best/.style={-{Triangle}, bend left=20, color=teal, line width=1pt},
    every loop/.style={min distance=12mm}
}

\colorlet{mygray}{gray!40}

\makeatletter
\let\oldnl\nl
\newcommand{\nonl}{\renewcommand{\nl}{\let\nl\oldnl}}
\makeatother

\makeatletter
\def\moverlay{\mathpalette\mov@rlay}
\def\mov@rlay#1#2{\leavevmode\vtop{
   \baselineskip\z@skip \lineskiplimit-\maxdimen
   \ialign{\hfil$\m@th#1##$\hfil\cr#2\crcr}}}
\newcommand{\charfusion}[3][\mathord]{
    #1{\ifx#1\mathop\vphantom{#2}\fi
        \mathpalette\mov@rlay{#2\cr#3}
      }
    \ifx#1\mathop\expandafter\displaylimits\fi}
\makeatother

\usepackage{thmtools,thm-restate}
\usepackage[capitalise,nameinlink,noabbrev]{cleveref}

\newtheorem{definition}{Definition}
\newtheorem{lemma}{Lemma}
\newtheorem{theorem}{Theorem}
\newtheorem{corollary}{Corollary}
\newtheorem{proposition}{Proposition}

\Crefname{claim}{Claim}{Claims}
\Crefname{corollary}{Corollary}{Corollaries}
\Crefname{definition}{Definition}{Definitions}
\Crefname{example}{Example}{Examples}
\Crefname{lemma}{Lemma}{Lemmas}
\Crefname{property}{Property}{Properties}
\Crefname{proposition}{Proposition}{Propositions}
\Crefname{remark}{Remark}{Remarks}
\Crefname{theorem}{Theorem}{Theorems}

\usepackage[linesnumbered,ruled,vlined]{algorithm2e}  

\theoremstyle{remark}

\newcommand{\bigcupdot}{\charfusion[\mathop]{\bigcup}{\cdot}}
\newcommand{\mes}{\mathsf{MES}}
\newcommand{\cG}{\mathcal{G}} 
\newcommand{\cA}{\mathcal{A}} 
\newcommand{\cN}{\mathcal{N}} 
\newcommand{\cV}{\mathcal{V}} 
\newcommand{\cL}{\mathcal{L}} 
\newcommand{\R}{\mathbb{R}_{{\ge}0}}

\let\displaystyle\textstyle

\title{(Almost Full) EFX for Three (and More) Types of Agents}

\author{
	\begin{tabular}{m{0.12\textwidth}m{0.12\textwidth}m{0.12\textwidth}m{0.12\textwidth}
 }
    \multicolumn{2}{c}{\textbf{Pratik Ghosal}} & \multicolumn{2}{c}{\textbf{Vishwa Prakash HV}\footnote{Supported by TCS Research Scholar Fellowship}}
        \\
        \multicolumn{2}{c}{Indian Institute of Technology Palakkad}\footnote{Part of this work was done when the author was at Chennai Mathematical Institute.} & \multicolumn{2}{c}{Chennai Mathematical Institute} 
        \\ 
		    \multicolumn{2}{c}{\href{mailto:pratik@iitpkd.ac.in}{\small{\texttt{pratik@iitpkd.ac.in}}}} &\multicolumn{2}{c}{\href{mailto:vishwa@cmi.ac.in}{\small{\texttt{vishwa@cmi.ac.in}}}}
        \\
        &&&\\
		\multicolumn{2}{c}{\textbf{Prajakta Nimbhorkar}} & \multicolumn{2}{c}{\textbf{Nithin Varma}} 
        \\
		\multicolumn{2}{c}{Chennai Mathematical Institute} & \multicolumn{2}{c}{University of Cologne, Germany}\footnote{Part of the work was done when the author was at Chennai Mathematical Institute and Max Planck Institute for Informatics, Saarland Informatics Campus, Germany.} 
        \\ 
		\multicolumn{2}{c}{\href{mailto:prajakta@cmi.ac.in}{\small{\texttt{prajakta@cmi.ac.in}}}} & \multicolumn{2}{c}{\href{mailto:nithvarma@gmail.com}{\small{\texttt{nithvarma@gmail.com}}}}
	\end{tabular}
}

\DeclareMathOperator*{\argmax}{arg\,max}
\DeclareMathOperator*{\argmin}{arg\,min}

\date{}

\begin{document}

\maketitle 
 

\input{sections/abstract}
\input{sections/intro}

\input{sections/prelims}
\input{sections/k_k-2}
\input{sections/n11}
\input{sections/conclusion}
\input{sections/ack}
\bibliographystyle{alphaurl}
\bibliography{references}

\end{document}

%% file: sections/abstract.tex
\begin{abstract}\label{sec:abstract}
    We study the problem of determining an envy-free allocation of indivisible goods among multiple agents with additive valuations. EFX, which stands for envy-freeness up to any good, is a well-studied relaxation of the envy-free allocation problem and has been shown to exist for specific scenarios. EFX is known to exist for three agents, and for any number of agents when there are only two types of valuations. EFX allocations are also known to exist for four agents with at most one good unallocated. 
    
    In this paper, we show that EFX exists with at most \(k-2\) goods unallocated for any number of agents having \(k\) distinct valuations. Additionally, we show that complete EFX allocations exist when all but two agents have identical valuations.
\end{abstract}

%% file: sections/intro.tex
\section{Introduction}\label{sec:intro}
Fair division of indivisible goods is a fundamental problem in the study of multi-agent systems. The problem concerns distributing indivisible resources fairly among agents. Several real-life scenarios reflect the importance of this problem. Examples include inheritance division, allocation of slots on computing machines to jobs etc. Maintaining fairness is challenging, especially when agents have heterogeneous preferences over subsets of items. Formally, the problem is to allocate a set $\cG = \{g_1, \dots g_m\}$ of $m$ goods to a set $\cA = \{a_1, \dots a_n\}$ of $n$ agents such that each good is allocated to at most one agent and each agent thinks of the overall allocation as being \emph{fair}. %

One of the most well-studied fairness notions is \emph{envy-freeness}. To formalize this notion, we model each agent $a_i$, $i \in [n]$, as having a valuation function $v_i: 2^\cG \to \R$ on subsets (\textit{bundles}) of goods. An allocation $X = (X_1, X_2, \dots X_n)$, where the bundle $X_i$ is allocated to agent $a_i$, $ i\in [n]$, is {\em envy-free} (EF) if each agent values their own bundle at least as much as that of any other agent, i.e., $v_i(X_i) \geq v_i(X_j)$ for all $i, j \in [n]$. 

EF allocations may not exist in general, the simplest instance being that of two agents and one valuable good. Various relaxations of envy-freeness have been proposed. The concept of \emph{envy-freeness up to one good} (EF1) was proposed in \cite{budish}.
An allocation $X$ is said to be EF1, if for each $i,j \in [n]$ there exists some good $g\in X_j$ such that $v_i(X_i)\geq v_i(X_j \setminus \{g\})$. 
EF1 allocations are known to exist and can be found in polynomial time~\cite{lipton}. 

In between the notions of EF and EF1 allocations, lies {\em envy-freeness up to any good} (EFX), introduced by \cite{efx_no1}.
Given an allocation $X$, an agent $a_i$ \emph{strongly envies} agent $a_j$ if there exists $g \in X_j$ such that $v_i(X_j \setminus \{g\})>v_i(X_i)$, i.e., $a_i$ envies the bundle $X_j$ of $a_j$ even after removing the good $g$ from $X_j$.  
The allocation is EFX if no agent strongly envies another agent. In other words, for each pair of agents $a_i$ and $a_j$, we have $v_i(X_i)\geq v_i(X_j \setminus \{g\})$ for any good $g\in X_j$. 

Unlike EF and EF1, the question of whether EFX allocations always exist is far from settled and is one of the important open questions in the field of discrete fair allocation. EFX allocations exist when all agents have the same valuation function, or when there are only two agents \cite{pr20}. In \cite{mahara,Mahara21}, the authors improved on this result and showed the existence of EFX for multiple agents when there are only two valuation functions. 
In a breakthrough result, \cite{efx_3} showed that EFX always exists for $3$ agents when the valuation functions of agents are additive.\footnote{A valuation $v: 2^\cG \to \R$ is additive if, for each bundle $S \subseteq \cG$ of goods, $v(S) = \sum_{g \in S} v(\{g\})$. The result of \cite{efx_3_simple}, which simplifies \cite{efx_3}, holds for more general valuation functions, which they call MMS-feasible valuations (see Definition~\ref{def:mms-feasible}).} %

All of the aforementioned results are for \emph{complete allocations} that allocate each good to some agent. %
In \cite{ChaudhuryKMS21}, the authors considered {\em incomplete allocations} that leave some goods unallocated and showed that an EFX allocation exists for $n$ agents when at most $n-1$ goods can remain unallocated, {while guaranteeing that no agent envies the bundle of unallocated goods}. This result was improved in \cite{BergerCFF22} where the authors show existence of EFX allocation for $n$ agents with at most $n-2$ unallocated goods, and also an EFX allocation for the special case of $4$ agents, with at most one unallocated good. {In both  cases, no agent envies the bundle of unallocated goods.}

\subsection{Our Contributions }\label{subsec:contributions}
In this work, we generalize the results of \cite{efx_3,efx_3_simple,mahara,BergerCFF22} to the case where there are multiple agents with the same valuation function. Our first result shows that EFX allocations with at most $k-2$ unallocated items always exist for $n$ agents when the valuation function of each agent is chosen from a set of $k$ distinct {\emph{nice-cancelable}\footnote{ \(v\) is said to be nice-cancelable if for any two bundles \(S,T\subseteq \cG\), and a good \(g\in \cG\setminus (S\cup T)\), \(v(S\cup g)>v(T\cup g)\implies v(S)>v(T).\)} valuation functions. In other words, among the $n$ agents, subsets of agents have identical valuation functions, so that collectively, there are at most $k$ distinct valuation functions. 
\begin{theorem}\label{thm:kk2}
    When there are $n$ agents such that the valuation of each agent is chosen from a set of $k$ distinct additive valuations, an EFX allocation exists that leaves at most $k-2$ goods unallocated. Furthermore, no agent envies the bundle of unallocated goods. {Moreover, this holds even when all the agents have \emph{nice-cancelable} valuations, a generalization of additive valuations.}
\end{theorem}
\begin{corollary}\label{cor:three-types}
    In an instance where each agent has one of three distinct nice-cancelable valuations, an EFX allocation with one unallocated good always exists.
\end{corollary}

Note that if we substitute \(k=2\) in Theorem~\ref{thm:kk2}, we get a complete EFX allocation for agents with two types of valuations. Therefore, Theorem~\ref{thm:kk2} generalizes the result of \cite{mahara_2}.

We also show a complete EFX allocation for a special case of Corollary~\ref{cor:three-types}, where among $n$ agents, $n-2$ have identical valuations and the other two agents can have different valuations.
We refer to this as the problem of EFX allocation when there are two \textit{outliers}. %

\begin{theorem}\label{thm:efx-n-agents}
 Consider a set of $n$ agents with additive valuations where at least $n-2$ agents have identical valuations. Then, for any set of goods, a complete EFX allocation always exists. Moreover, this holds even when all the agents have MMS-feasible valuations, a generalization of additive valuations. %
\end{theorem}

This generalizes \cite{efx_3_simple} and \cite{efx_3} as it implies EFX for $n$ agents when \(n=3\).

\subsection{Overview of our Techniques}

{
Here, we describe at a high level, the key ideas involved in the proofs of our main theorems.}

For proving Theorem~\ref{thm:kk2},
we consider \(k\) groups of agents such that all agents in a given group have identical valuations. The agent with the least valuable bundle in a group is called the \emph{leading agent} of that group. We start with an EFX allocation where each agent gets at most one good. We then iteratively find a new EFX allocation in which Pareto dominates the previous ones until we find our desired EFX allocation. To achieve this, we focus on the set $\cL$ of leading agents in each group. To improve a given EFX allocation \(X\), we consider the sub-allocation \(X({\cL})\) restricted to the leading agents and replace \(X({\cL})\) with a Pareto dominating EFX allocation (restricted to the leading agents) \(Y({\cL})\). However, not all such Pareto dominating sub-allocations extend to an allocation that is EFX for all the agents, as some of the non-leading agents may strongly envy the new bundles of the leading agents. To address this problem, in Lemma~\ref{lemma:replacing_core}, we show that if we replace \(X({\cL})\) with a special Pareto dominating EFX allocation \(Y({\cL})\), where every leading agent gets a \emph{minimally envied subset} with respect to their previous bundle, then replacing \(X({\cL})\) with such a \(Y({\cL})\) gives an allocation that is EFX for all the agents. We observe that when the allocation \(X(\cL)\) is envy-free, a result in \cite{BergerCFF22} can be used to compute such a special allocation \(Y({\cL})\). On the other hand, if the allocation \(X(\cL)\) is not envy free, we observe that there can at most be \(k-1\) sources in the envy graph (discussed in Proposition~\ref{prop:sources}) and therefore, a result in \cite{ChaudhuryKMS21} can be used to get a Pareto dominating allocation in this situation. 
}

Our proof of Theorem~\ref{thm:efx-n-agents} begins by considering an {\em almost feasible EFX allocation} ({see Definition~\ref{def:almost-feasible}})  consisting of $n$ bundles. An almost EFX feasible allocation ensures that \(n-1\) bundles are all EFX feasible {(see Definition~\ref{def:EFX-feasible})} for each of the \(n-2\) agents with identical valuations and that the remaining bundle is EFX feasible for one of the two outlier agents. 
Our procedure modifies such an allocation carefully to either get to an EFX allocation, in which case we are done, or to another almost EFX feasible allocation. The termination of our procedure is ensured by the fact that the resulting almost EFX feasible allocation is strictly better than the previous one in a concrete sense, i.e., in terms of a potential function {that cannot grow forever}.
The challenge lies in maintaining the aforementioned invariant and proving the increase in potential. 

\subsection{Related Work}    
The notion of envy-free allocations was introduced by \cite{gamow1958puzzle} and \cite{foley1967resource}. For indivisible goods, \cite{lipton} and \cite{budish} consider a relaxed notion of envy-freeness known as {\em envy-freeness up to one good (EF1)}. The notion of envy-freeness up to any good (EFX) was introduced by \cite{efx_no1}. The existence of EFX allocations has been shown in various restricted settings like $2$ agents with arbitrary valuations and any number of agents with identical valuations \cite{pr20}, for additive valuations with $3$ agents \cite{efx_3}, at most two valuations for an arbitrary number of agents \cite{mahara, Mahara21}, for the case when each value of each agent can take one of the two possible values \cite{AmanatidisBFHV21}, etc. EFX allocations for the case when some goods can be left unallocated have been considered in several works \cite{BramsKK22,ColeGG13,CaragiannisGH19} etc. \cite{CaragiannisGH19} show that discarding some items can achieve at least half of the maximum Nash Welfare whereas \cite{ChaudhuryKMS21} show that an EFX allocation always exists for $n$ agents with arbitrary valuations with at most $n-1$ unallocated items, \cite{BergerCFF22} improve this to {show the existence of EFX with at most \(n-2\) goods unallocated, and} for $4$ agents with at most one unallocated good. For further works related to to fair allocation, we refer the reader to a recent survey by \cite{amanatidis2023fair}.

%% file: sections/prelims.tex
\section{Preliminaries}\label{sec:prelim}
\sloppy
Let \(\cA=\{a_1,a_2,\cdots,a_n\}\) be a set of \(n\) agents and let \(\cG=\{g_1,g_2,\cdots,g_m\}\) be a set of \(m\) indivisible goods. An instance of discrete fair division is specified by the tuple \(\langle \cA,\cG, \cV \rangle\), where \(\cV = \{v_1(\cdot), v_2(\cdot),\cdots,v_n(\cdot)\}\) is such that for $i \in [n]$, the function \(v_i: 2^{\cG} \to \R\) denotes the valuation of agent \(a_i\) on subsets of goods. 
\fussy
\sloppy We use the term \emph{bundle} to denote a subset of goods. For bundles $X_1, \ldots,  X_k$, and an agent $a$, we use $\max_a (X_1, \ldots, X_k)$ (resp. $\min_a (X_1, \ldots, X_k)$) to denote the bundle that is most (resp. least) valued as per the valuation function $v_a$. 
An \emph{allocation} is a tuple $X = \langle X_1, X_2, \dots, X_n \rangle$ of $n$ mutually disjoint bundles such that bundle $X_i$ is assigned to agent $a_i$ for all $i \in [n]$. 
Given an allocation \(X=\langle X_1, X_2, \cdots,X_n \rangle\), we say that agent $a_i$ \emph{envies} another agent $a_j$ if $v_i(X_j) > v_i(X_i)$. We often also say that agent \(a_i\) \emph{envies the bundle} \(X_j\).  
\sloppy
\begin{definition}[Strong envy]
Given an allocation \(X=\langle X_1, X_2, \cdots,X_n \rangle\), an agent \(a_i\) \emph{strongly envies} an agent \(a_j\) if \( v_i(X_j\setminus \{g\}) > v_i(X_i)\) for some \(g\in X_j\).    
\end{definition}
\fussy
An allocation is EFX if there is no strong envy between any pair of agents.
\begin{definition}[EFX-feasibility]\label{def:EFX-feasible} A bundle \(S\subseteq \cG\) is said to be EFX-feasible \emph{w.r.t.} a disjoint bundle \(T\) according to valuation \(v\), if for all \(h\in T\), \(v(T\setminus \{h\}) \le v(S)\). 
Given an allocation \(X=\langle X_1,X_2,\cdots,X_n \rangle\), bundle \(X_i\) is EFX-feasible \emph{for an agent} \(a_j\) if \(X_i\) is EFX-feasible w.r.t. all other bundles in \(X\) according to valuation \(v_j\).
\end{definition}
An allocation \(X=\langle X_1,X_2,\cdots,X_n\rangle\) is said to be EFX if for all $i \in [n]$, the bundle \(X_i\) is EFX-feasible for agent $a_i$.
Let $a \in \cA; g \in \cG;\ \ S,T \subseteq \cG \text{ and } v: 2^\cG \to \R$. To simplify notation, we write \(v(g)\) to denote \(v(\{g\})\) and use \(S\setminus g\), \(S\cup g\) to denote \(S\setminus\{g\}\), \(S\cup\{g\}\), respectively. We also write \(S>_a T\) to denote \(v_a(S)>v_a(T)\) and similarly for \(<_a, \ge{_a},\le{_a} \) and \(=_a \). We use \(\min_a(S,T)\) and \(\max_a(S,T)\) to denote \(\argmin_{Y\in\{S,T\}} v_a(Y)\) and \(\argmax_{Y\in\{S,T\}} v_a(Y)\). 
\begin{definition}[Minimally envied subset \cite{ChaudhuryKMS21}] Given an allocation \(X\), if an agent \(a_i\) with bundle \(X_i\) envies a bundle \(S\), we call \(T\subseteq S\) a \emph{minimally envied subset} of \(S\) for agent \(a_i\) if both the following conditions hold.
\begin{enumerate}
    \item \(v_i(X_i) < v_i(T)\)
    \item \(v_i(X_i) \ge v_i(T\setminus h)~~\forall {h\in T}\)
\end{enumerate}
\end{definition}
We generalize the above definition to a minimally envied subset with respect to a subset of agents.
\begin{definition}\label{def:general_mes}
    Let \(N \subseteq \cA\) be a non-empty set of agents, \(X(N)\) be an allocation restricted to the agents in \(N\), and \(S \subseteq \cG\) be  a set of goods such that at least one agent in \(N\) envies \(S\).
    
    We call \(T\subseteq S\) as the \emph{minimally envied subset} of \(S\) with respect to \(N\) and \(X(N)\), denoted by \(\mathsf{MES}_{N}(X(N),S)\),  if at least one agent in \(N\) envies \(T\) and no agent in \(N\) strongly envies \(T\). We call the set of agents in \(N\) who envy \(T\) as the \emph{most envious agents} of \(S\) among \(N\). Further, \(S\setminus T\) is refereed to as a \emph{discard} set. 
\end{definition}
Note that \(\mes_{N}(X(N),S)\) need not be unique. However, the size of the minimally envied subset \(|\mes_{N}(X(N),S)|\) is fixed for a given \(N,X(N)\) and \(S\).
\begin{definition}
    Given an allocation \(X\) and a subset \(S\subseteq \cG\) of goods such that at least one agent envies \(S\), we say that agent \(a_i\) \emph{champions} \(S\) if \(a_i\) envies \(\mathsf{MES}_{\cA}(X,S)\).
\end{definition}
Given an allocation \(X\), the envy relation among the agents is graphically represented as follows. 
\begin{definition}[Envy Graph]
    Given an allocation \(X\), let \(E_X=(\cA,E)\) be a directed graph where the set of agents \(\cA\) is the set of vertices and for every pair \((a_i,a_j)\) of agents, \((a_i,a_j)\in E\) iff agent \(a_i\) envies agent \(a_j\) under the allocation \(X\).
\end{definition}
\sloppy \begin{definition}
    An allocation \(X\) is said to Pareto dominate another allocation \(Y\),denoted by \(X\succ Y\), if for every agent \(a_i\in \cA\), \(v_i(X_i)\ge v_i(Y_i)\), and there exists at least one agent \(a_j\) such that \(v_j(X_i)>v_j(Y_i)\).
\end{definition}
{Given an allocation \(X\) and its envy graph \(E_X\), an agent \(a\) is said to be a source in \(E_X\) if no agent envies \(a\). When there are few source agents and sufficiently many unallocated goods, we use the following result by \cite{ChaudhuryKMS21} to show Pareto improvement:}
\begin{lemma}[\cite{ChaudhuryKMS21}]\label{lemma:less_source_more_goods}
    Let \(X\) be a partial EFX allocation with at least \(k\) unallocated goods. If the envy graph \(E_X\) has at most \(k\) source vertices, then there exists another EFX allocation \(Y\) such that \(Y\succ X\).
\end{lemma}
\sloppy \paragraph*{\textbf{Non-degenerate instances \cite{efx_3,efx_3_simple}}} An instance \( \mathcal{I} =\langle \cA,\cG,\cV \rangle\) is said to be \emph{non-degenerate} if and only if no agent values two different bundles equally. That is, \(\forall{a_i \in \cA}\) we have \(v_i(S) \ne v_i(T)\) for all \(S\ne T,\text{ where } S,T\subseteq \cG\). \cite{efx_3_simple} showed that it suffices to deal with non-degenerate instances when there are \(n\) agents with general valuation functions, i.e., if each non-degenerate instance has an EFX allocation, each general instance has an EFX allocation.
 In the rest of the paper, we only consider non-degenerate instances. This implies that all goods are positively valued by all agents as value of the empty bundle is assumed to be zero. 
Non-degenerate instances have the following property.
\begin{proposition}\label{prop:identical_envy}
Given any allocation \(X\), for any two agents \(a\) and \(a'\) with identical valuations, either \(a\) envies \(a'\) or vice versa.
\end{proposition} 
\paragraph*{\textbf{Properties of valuation functions}}
A valuation \(v\) is said to be \emph{monotone} if \(S\subseteq T\) implies \( v(S) \leq v(T) \) for all \(S,T \subseteq \cG\). 
Monotonicity is a natural restriction on valuation functions and occurs frequently in real-world instances of fair division. 
A valuation \(v\) is \emph{additive} if \(v(S) = \sum_{g\in S} v(\{g\})\) for all \(S\subseteq \cG\). Additive valuation functions are, by definition, also monotone. 
{A generalization of additive valuations are \emph{nice cancelable valuations}, introduced by \cite{BergerCFF22}.}
{
\begin{definition}\label{def:nice_cancelable}
    A valuation function \(v\) is said to be \emph{nice cancelable} if \(v\) is non-degenerate and for any two bundles \(S,T\subseteq \cG\), and a good \(g\in \cG\setminus (S\cup T)\),
    \[v(S\cup g)>v(T\cup g)\implies v(S)>v(T).\]
\end{definition}}
{
Nice cancelable valuations include \emph{budget additive} \((v(S)=\min(\sum_{g\in S}v(g),c))\), \emph{unit demand} \((v(S)=\max_{g\in S}v(g))\) and \emph{multiplicative} \((v(S)=\prod_{g\in S} v(g))\) valuations~\cite{BergerCFF22}.}
{
\cite{efx_3_simple} introduced a more general class of valuation functions called as the MMS-feasible valuations.}
\begin{definition}\label{def:mms-feasible}
A valuation \(v:2^{\cG}\to\R\) is MMS-feasible if for every subset of goods \(S\subseteq \cG\) and     every pair \(A=(A_1,A_2)\), \(B=(B_1,B_2)\) of partitions of \(S\), we have
    \[\max(v(B_1),v(B_2)) > \min(v(A_1),v(A_2)).\]
\end{definition}
\paragraph*{\textbf{Algorithm of Plaut and Roughgarden}} For monotone valuation functions, \cite{pr20} gave an algorithm to compute an EFX-allocation when all agents additionally have the same valuation \(v(\cdot)\) function.
Throughout this paper, we refer to this algorithm as the PR~algorithm. Let \(M\subseteq\cG\) be a subset of goods. 
Let \(X=\{X_1,X_2,\cdots,X_k\}\) be a \(k\)-partition of \(M\). In its most general form, the PR algorithm takes \((X,v,k)\) as input and outputs a (possibly different) $k$-partition \(Y=\{Y_1,Y_2,\cdots,Y_k\}\) of $M$.
We crucially use the following properties~\cite{pr20} of the output of the PR algorithm. 
\begin{enumerate}
    \item For all \(i\in[n]\), if \(Y_i\) is allocated to agent \(a\) then agent \(a\) does not strongly envy any other bundle in \(Y\). 
    \sloppy \item The value of the least valued bundle does not decrease, i.e., \(\min(v(Y_1),v(Y_2),\cdots,v(Y_k)) \ge \min(v(X_1),v(X_2),\cdots,v(X_k)).\) 
\end{enumerate}

%% file: sections/k_k-2.tex
\section{Almost EFX for \(k\) Types of Agents}\label{sec:k_types}
 
We prove Theorem~\ref{thm:kk2} in this section. Recall that the input instance consists of a set of agents \(\cA\) that can be partitioned as \(\cA = \bigcupdot_{i=1}^k \cA_i\), such that all the agents of any given part \(\cA_i\) have an identical valuation, say \(v_i\). We refer to this setting as the one with \(n\) agents and \(k\) \textit{types} of valuations. Here \(v_1, v_2, \ldots, v_k\) represent the \(k\) different valuation functions. 

For all \(i\in[k]\), let \(|\cA_i|=n_i\).
Let \(\mathcal{A}_i = \{a_1^i, a_2^i, \ldots, a_{n_i}^i\}\) be the set (or group) of agents with valuation function \(v_i\). 
Given an allocation \(X\), let \(X_j^i\) represent the bundle allocated to agent \(a_j^i\). In any given group \(\cA_i\), we can assume w.l.o.g. that agent \(a_1^i\) gets the least valued bundle, followed by agent \(a_2^i\) and so on. In other words, the bundle $X_j^i$ is the $j$-th smallest bundle in group $\cA_i$ in the allocation $X$. For any given group \(\cA_i\), we call the agent \(a_1^i\) as the \emph{leading agent} of that group. 

We now make a few essential observations about the envy graph \(E_X\) of any given allocation \(X\) for an instance with \(k\)-types of agents. 

\begin{proposition}\label{prop:sources}
     Given an instance with \(k\)-types of agents, and an allocation \(X\), a non-leading agent can never be a source vertex in the envy graph \(E_X\). Hence $E_X$ has at most $k$ sources.
 \end{proposition}
 
 This is due to the fact that every non-leading agent is envied by the leading agent of the same group. Similarly, we have the following observation:

 \begin{proposition}\label{prop:if_someone_envies_leading_agent_envies}
     If an agent \(a_p^i\) of group \(\cA_i\) envies (or strongly envies) an agent \(a\), then the leading agent \(a_1^i\) also envies (respectively strongly envies) agent \(a\).
 \end{proposition}

Let \(\cL=\{a_1^1,a_1^2,\ldots,a_1^k\}\) be the set of leading agents of each group. Given an allocation \(X\), let \(S\subseteq \cG\) be a bundle of goods that some agent \(a\in \cA\) envies. Then, we have the following observations: 

\begin{proposition}\label{prop:MES_projection}
    If \(T=\mes_{\cL}(X(\cL),S)\) is a minimally envied subset of \(S\) with respect to \(\cL,X(\cL)\), then \(T\) is also a minimally envied subset with respect to \(\cA,X\). That is, \(T=\mes_{\cA}(X,S)\).
\end{proposition}
\begin{proof}
    For the sake of contradiction, assume that \(T\ne\mes_{\cA}(X,S)\). That is, some agent \(a_j^i\) strongly envies a strict subset \(R\subsetneq T\). Then, from Proposition~\ref{prop:if_someone_envies_leading_agent_envies}, we know that the leading agent \(a_1^i\) also envies \(R\). This contradicts our assumption that \(T=\mes_{\cL}(X(\cL),S)\).
\end{proof}

We now prove a lemma that will be used crucially to prove Theorem~\ref{thm:kk2}. In the lemma below, we assume that there is an existing (possibly partial) EFX allocation $X$. We show that if we improve the bundles of the leading agents such that the new bundle of each leading agent is a minimally envied subset with respect to their respective bundles in $X$, then the resulting allocation is EFX for all agents.
\begin{lemma}\label{lemma:replacing_core}
    Let \(X\) be an EFX allocation (not necessarily complete), and \(X(\cL)\) be the allocation \(X\) restricted to the set of leading agents \(\cL\). Let \(Y(\cL)=\langle Y_1^1,Y_1^2,\ldots,Y_1^k\rangle\) be a new allocation for the agents in \(\cL\) such that \(Y(\cL)\succ X(\cL)\) and $Y(\cL)$ is an EFX allocation within $\cL$. Moreover, \(\forall i\in k\), \(\exists S\subseteq\cG,\) such that \(Y_1^i=\mes_{\cL}(X(\cL),S)\). Then, \(X'=  X({\cA\setminus \cL}\cup Y(\cL))\) is an EFX allocation for \(\cA\) and \(X'\succ X\).
\end{lemma}

\begin{proof}
    As no agent gets a worse bundle in \(X'\) as compared to \(X\), and at least one agent (happens to be in \(\cL\)) is strictly better off, \(X'\) Pareto dominates \(X\). It remains to be shown that $X'$ is an EFX allocation.

    First we show that the leading agents do not strongly envy anyone in $X'$. Consider an arbitrary leading agent \(a_1^j\). Agent \(a_1^j\) does not strongly envy any other leading agent in \(X'\) as \(Y(\cL)\) is an EFX allocation within $\cL$. Agent \(a_1^j\) does not strongly envy any non-leading agent \(a_p^q\) since \(v_j(Y_1^j)\ge v_j(X_1^j)>v_j(X_p^q\setminus h)\), \(\forall h\in X_p^q\).
    
    Now we show that the non-leading agents do not strongly envy anyone. Consider an arbitrary non-leading agent \(a_p^q\). Agent \(a_p^q\) does not strongly envy any other non-leading agent as they both retain the same bundles as in $X$. We know that in \(X'\), every leading agent holds a minimally envied subset of the form \(\mes_{\cL}(X(\cL),S)\). From Proposition~\ref{prop:MES_projection}, we know that \(\mes_{\cL}(X(\cL),S) = \mes_{\cA}(X,S)\). Therefore, agent \(a_p^q\) does not strongly envy any leading agent either. This proves that \(X'\) is an EFX allocation.
\end{proof}

\paragraph{Proof outline for Theorem~\ref{thm:kk2}:}
The proof of Theorem~\ref{thm:kk2} involves two cases. Let $X$ be a (possibly partial) EFX allocation. By Proposition~\ref{prop:sources}, the envy graph $E_X$ has at most $k$ sources. When the number of unallocated goods is at least as large as the number of sources in $E_X$, by Lemma~\ref{lemma:less_source_more_goods}, there is another EFX allocation $Y$ such that $Y\succ X$. Thus there is an EFX allocation with at most $k-1$ unallocated goods. Now we adapt the technique from \cite{BergerCFF22} to get an EFX allocation with at most $k-2$ unallocated goods.

\fussy Consider the case when there are only \(k\) agents with possibly distinct valuations. {Let \(\cN\) be the set of these agents.} 
Then an EFX allocation with at most \(k-2\) unallocated goods is known to exist due to \cite{BergerCFF22}. If the envy graph has $k-1$ or fewer sources, an EFX allocation with at most $k-2$ unallocated goods can be obtained from \cite{ChaudhuryKMS21}. It remains to obtain the same when the envy graph has $k$ sources. Thus there is no envy among the $k$ agents, and hence $X$ is a (possibly partial) envy-free allocation. In Lemma~4.2 of \cite{BergerCFF22} below, when given an \emph{envy-free} allocation for $k$ agents, with exactly \(k-1\) unallocated goods, they find a new EFX allocation \(Y\), such that \(Y\succ X\). We observe that \(Y\) has a special property: in \(Y\), every agent receives a bundle of the form \(Y_i = \mes_{\cN}(X, S)\), where \(S\) is either \(X_j \cup g\) or \(X_j \cup D\), with \(D\) being some discard set.

\begin{lemma}[\cite{BergerCFF22}]\label{lemma:berger}
    Let \(\cN\) consist of exactly \(k\) agents each with possibly different valuation. Now, given a (possibly partial) envy-free allocation \(X\) with exactly \(k-1\) goods unallocated, there exists an EFX allocation \(Y\) (possibly partial) such that \(Y\succ X\). Moreover, in \(Y\), every agent receives a bundle of the form \(Y_i = \mes_{\cN}(X, S)\), where \(S\) is either \(X_j \cup g\) or \(X_j \cup D\), {for some bundle \(X_j\), unallocated good \(g\), and \(D\) being some discard set.}
\end{lemma}

We now prove the following two lemmas, which will be used as subroutines to give a constructive proof of Theorem~\ref{thm:kk2}.

\begin{lemma}\label{lemma:main}
    Consider an instance with \(k\)-types of agents. If there exists an EFX allocation \(X\) with at least \(k-1\) unallocated goods, then there exists a EFX allocation (not necessarily complete) \(Y\), such that \(Y\succ X\).
\end{lemma}
\begin{proof}
    Consider the envy graph \(E_X\) of of the allocation \(X\). From Proposition~\ref{prop:sources}, we know that only the leading agents can be the sources. Therefore, there can be at most \(k\) sources. If the number of sources is less than or equal to \(k-1\), since there are at least \(k-1\) unallocated goods in \(X\), by applying Lemma~\ref{lemma:less_source_more_goods} we get an EFX allocation \(Y\), \(Y\succ X\). 

    Now, consider the case when there are exactly \(k\) sources in \(E_X\). From Proposition~\ref{prop:sources}, we know that the set of leading agents, \(\cL\) is exactly the set of sources in \(E_X\). If we consider the sub-allocation \(X(\cL)\) restricted to \(\cL\), observe that \(X(\cL)\) is envy free. We now apply Lemma~\ref{lemma:berger} on \(X(\cL)\) to get an EFX allocation \(Y(\cL)\) restricted to \(\cL\), such that \(Y(\cL)\succ X(\cL)\), and each agent in \(\cL\) gets a bundle of the form \(Y_1^i=\mes_{\cL}(X(\cL),S)\). Due to Lemma~\ref{lemma:replacing_core}, the allocation \(X({\cA\setminus \cL}\cup Y(\cL))\) Pareto dominates \(X\). 
\end{proof}

\begin{lemma}\label{lemma:no_envy_towards_unallocated}
    Let \(X\) be a partial EFX allocation, and \(U\ne\varnothing\) be the set of unallocated goods. If some agent \(a\) envies \(U\), then there exists an EFX allocation (not necessarily complete) \(Y\), such that \(Y\succ X\).
\end{lemma}
\begin{proof}
    If some agent \(a\) envies \(U\), then there exists a champion agent \(a'\) who envies the minimally envied subset \(T=\mes_{\cN}(X,U)\). We replace the bundle of agent \(a'\) with \(T\). This gives a new allocation \(Y\) which Pareto dominates \(X\).
\end{proof}

\begin{proof}[Proof of Theorem~\ref{thm:kk2}]
    Consider an instance with \(k\)-types of agents. If the number of goods, \(|\cG|\le n\), then any allocation with at most one good per agent is an EFX allocation. So assume $|\cG|>n$. We begin with a partial, trivially EFX allocation \(X\) where each agent has exactly one good. Let \(U\) be the set of unallocated goods.

    As long as \(|U|\ge k-1\) or some agent envies \(U\), we find a new allocation \(Y\) that Pareto dominates \(X\) using Lemma~\ref{lemma:main} and Lemma~\ref{lemma:no_envy_towards_unallocated}. Since there are finitely many EFX allocations, this procedure must end with an EFX allocation with at most \(k-2\) unallocated goods and no agent envying \(U\).%
\end{proof}

%% file: sections/n11.tex
\section{EFX with Two Outlier Valuations}\label{sec:n11-efx}

In this section, we show that EFX allocation always exists for \(n\) agents when \(n-2\) of the agents have identical valuations thus prove~Theorem~\ref{thm:efx-n-agents}. 

Consider a set of \(n\) agents \(\cA = \{a_1, a_2,\cdots,a_{n-2}, b_1, c_1\}\), a set of \(m\) goods \(\cG = \{g_1,g_2,\cdots,g_m\}\) and a set of three valuation functions \(\cV = \{v_a,v_b,v_c\}\) such that the agents \(a_1,a_2,\cdots,a_{n-2}\) have valuation \(v_a\) and outlier agents \(b_1\) and \(c_1\) have valuations \(v_b\) and \(v_c\) respectively. The valuations \(v_a\) and \(v_b\) are assumed to be monotone and \(v_c\) is assumed to be MMS-feasible. 

{\setlength\emergencystretch{\hsize}
\begin{definition}\label{def:almost-feasible}
We say that an allocation \(X = \langle X_1,X_2,\cdots,X_n \rangle\) is \emph{almost EFX-feasible} if it satisfies the following conditions: 
  \begin{enumerate}
  \item The first \(n-1\) bundles \(X_1,X_2,\cdots,X_{n-1}\) are EFX-feasible for agents~\(a_1,a_2,\cdots,a_{n-2}\).
  \item \(X_n\) is EFX-feasible for at least one outlier (that is either agent \(b_1\) or agent \(c_1\)).
  \end{enumerate}
  \label{def:invariants}
\end{definition}}

We define a potential function \(\phi\) which assigns a real value for each allocation \(X = \langle X_1,X_2,\cdots,X_n \rangle\) as follows:
\[
  \phi(X) = \min\{v_a(X_1),v_a(X_2),\cdots,v_a(X_{n-1})\}.
\]

\sloppy To prove Theorem~\ref{thm:efx-n-agents}, we first show that almost EFX-feasible allocations always exist. Then we show that, if an allocation \(X\) is almost EFX-feasible, then either \(X\) is an EFX allocation or there exists another almost EFX-feasible allocation \(X'\) with a strictly higher potential value, i.e., \(\phi(X')>\phi(X)\).  Since \(\phi(X)\) cannot grow arbitrarily as \(\phi(X)<v_a(\cG)\), there must exist an almost EFX-feasible allocation which is also an EFX allocation.

\sloppy \begin{proof}[Proof of ~Theorem~\ref{thm:efx-n-agents}:]
  
  For any given instance with \(n\) agents such that \(n-2\) agents have identical valuations, an almost EFX-feasible allocation always exists. This can be obtained by running the PR algorithm on \(\cG\) with the valuation \(v_a\) for all $n$ agents. Lets call this initial  allocation \(X = \langle X_1,X_2,\cdots,X_n\rangle\). From Property~1 of the PR algorithm, all the bundles are EFX-feasible for agents \(a_1,a_2,\cdots,a_{n-2}\). Let agent \(c_1\) pick the most valued bundle from \(X\) according their valuation \(v_c\). Without loss of generality, we can assume that the bundle picked by agent \(c_1\) is \(X_n\). It is clear that \(X_n\) is EFX-feasible for \(c_1\). Hence \(X\) is almost EFX-feasible.\\

  If either one among the agents \(b_1\) or \(c_1\) has at least one EFX-feasible bundle other than \(X_n\), say \(X_k\), then we are done. We allocate \(\langle X_n,X_k\rangle\) to agent \(c_1\) and \(b_1\) respectively, and the remaining bundles to agents \(a_1,a_2,\cdots,a_{n-2}\) arbitrarily. The resulting allocation is EFX.\\

 In the remainder of the proof, we consider the case that \(X_n\) is the only EFX-feasible bundle for both \(b_1\) and \(c_1\).

  Let \(g_b\) and \(g_c\) be the least valuable good(s) in \(X_n\) according to agents \(b_1\) and \(c_1\), respectively. Since \(X_n\) is the most valued bundle and also the \emph{only} EFX-feasible bundle in $X$ for agent \(b_1\) (or \(c_1\)), even if we give the maximum valued bundle from \(\{X_1,X_2,\cdots,X_{n-1}\}\) according to $v_b$ ($v_c$, respectively) to agent \(b_1\) (\(c_1\), respectively), they would strongly envy the bundle \(X_n\). That is
  \begin{equation}
    X_n\setminus g_b >_b \max_b(X_1,X_2,\cdots,X_{n-1})\label{eq:1}
  \end{equation}
  \begin{equation}
    X_n\setminus g_c >_c \max_c(X_1,X_2,\cdots,X_{n-1})\label{eq:2}
  \end{equation}

  Without loss of generality, assume
  \begin{equation}
    X_1<_a X_2 <_a \cdots <_a X_{n-1}\label{eq:3}
  \end{equation}

  \noindent Now, we consider the cases which arise when we move the least valued good from \(X_n\) (according to \(b_1\) or \(c_1\)) and add it to the bundle \(X_1\).

   \textbf{Case 1:} The bundle \(X_n\setminus g_b\) remains to be the most favorite bundle for agent \(b_1\) \emph{or} the bundle \(X_n\setminus g_c\) remains to be the most favorite bundle for agent \(c_1\). That is, 
    \begin{align*}
      &X_n\setminus g_b >_b X_1 \cup g_b, \text{ or }
      &X_n\setminus g_c >_c X_1 \cup g_c
    \end{align*}

    Assume that \(X_n\setminus g_b >_b X_1 \cup g_b\). The procedure is analogous if we consider the case that \(X_n\setminus g_c >_c X_1 \cup g_c\). 
    The new allocation is \( X' = \langle X_1\cup g_b,X_2,\cdots,X_n\setminus g_b \rangle\). Combining \(X_n\setminus g_b >_b X_1 \cup g_b\) with ~Equation~\ref{eq:1}, we get that the bundle \(X_n\setminus g_b\) is the most valuable according to $v_b$ and hence EFX-feasible for agent \(b_1\) in the new allocation.\\

    \textbf{Case 1.1: }\label{case:1.1} \(X_1 \cup{g_b} <_a X_2\).\\
      Combining \(X_1\cup{g_b} >_a X_1\) and~Equation~\ref{eq:3}, we can see that \[\phi(X')=v_a(X_1\cup{g_b})>v_a(X_1)=\phi(X).\] Thus there is an increase in the potential. For agents \(a_1,a_2,\cdots,a_{n-2}\), the bundle \(X_1 \cup {g_b}\) remains EFX-feasible as no other bundle has increased in value. Furthermore, for agents \(a_1,a_2,\cdots,a_{n-2}\), the bundles \(X_2,X_3,\cdots,X_{n-1}\) are EFX-feasible when compared to \(X_1\cup g_b\) as they are more valuable than \(X_1 \cup g_b\) according to $v_a$. They are also EFX-feasible when compared to \(X_n\setminus g_b\) because they were EFX-feasible against a higher valued bundle \(X_n\). Thus, bundles \(X_1 \cup g_b, X_2,\cdots,X_{n-1}\) are EFX-feasible for agents \(a_1,a_2,\cdots,a_{n-2}\). Therefore, the new allocation is almost EFX-feasible and has an increased potential.\\
      
     \textbf{Case 1.2}\footnote{Note that we do not have to consider the case that $X_1 \cup g_b =_a X_2$ since the instance is assumed to be non-degenerate.}: \(X_1 \cup{g_b} >_a X_2\).\\
      Let \((X_1\cup g_b) \setminus Z\) be a \emph{minimally envied subset} with respect to \(X_2\) under valuation \(v_a\). That is,
      \begin{equation}\label{eq:4}
        \begin{split}
          (X_1 \cup {g_b})\setminus Z >_a X_2 \ , and\\
          ((X_1 \cup{g_b})\setminus Z) \setminus {h} <_a X_2&\ \ \ \forall h\in (X_1 \cup{g_b})\setminus Z
        \end{split}
      \end{equation}
      Now, let the new allocation be
      \begin{align*}
        X' &= \langle X_1',X_2',\cdots,X_n'\rangle\\
           &= \langle (X_1 \cup{g_b})\setminus Z,\ X_2,\cdots ,(X_n\setminus \{g_b\})\cup Z \rangle\\
      \end{align*}

      Since \((X_1 \cup{g_b})\setminus Z >_a X_2\), $\phi(X')=v_a(X_2) >v_a({X_1})=\phi(X)$, and thus the potential has strictly increased.
            From ~Equation~\ref{eq:1}, we have \(X_n\setminus g_b >_b \max_b(X_1,X_2,\cdots,X_{n-1})\). From the Case~1 assumption, we also have \(X_n \setminus {g_b} >_b X_1\cup{g_b}\). Therefore, \[X_n' = (X_n \setminus {g_b})\cup Z >_b \max_b(X_1',X_2',\ldots,X_{n-1}')\]Thus \(X_n'\) is EFX-feasible for agent \(b_1\).
      
      \sloppy Next, we show that the bundles \(X_1',X_2',\cdots,X_{n-1}'\) are EFX-feasible \emph{among themselves} (i.e, not compared with \(X_n'\)) to agents \(a_1,a_2,\cdots,a_{n-2}\).\\
      The bundle \(X_1\) was EFX-feasible \emph{w.r.t.} \(X_2,\cdots,X_{n-1}\) in \(X\). Therefore, \(X_1'>_a{X_1}\) is also EFX-feasible \emph{w.r.t.} \(X_2',\cdots,X_{n-1}'\).\\
      Bundles \(X_2',\cdots,X_{n-1}'\) are EFX-feasible \emph{w.r.t.} each other as they remain unchanged. From ~Equation~\ref{eq:4} we know that \(X_1'\setminus{h} = ((X_1 \cup{g_b})\setminus Z) \setminus {h} <_a X_2\ \forall h\in ((X_1 \cup{g_b})\setminus Z)\), and from ~Equation~\ref{eq:3} we have \(X_2 <_a \cdots <_a X_{n-1}\). Therefore, both \(X_2',\cdots,X_{n-1}'\) are EFX-feasible \emph{w.r.t.} \(X_1'\) for agents \(a_1,a_2,\cdots,a_{n-2}\). Therefore, the bundles \(X_1',X_2',\cdots,X_{n-1}'\) are EFX-feasible among themselves to agents \(a_1,a_2,\cdots,a_{n-2}\).

      All that remains is to check the EFX-feasibility of bundles \(X_1',X_2' ,\cdots,X_{n-1}'\) \emph{w.r.t.} \(X_n'\). If the bundles \(X_1',X_2',\cdots, X_{n-1}'\) are EFX-feasible \emph{w.r.t.} \(X_n'\), then we meet all the conditions of the invariant and hence \(X'\) is almost EFX-feasible. Since $\phi(X') > \phi(X)$, we have an almost EFX-feasible solution with increased potential and we are done.


      Now, consider the case that one of the bundles in \(\{X_1',X_2',\cdots,X_{n-1}'\}\) is \emph{not} EFX-feasible \emph{w.r.t.} \(X_n'\). That is,

      \begin{align*}
        &\exists h\in X_n'~~\text{such that } X_n' \setminus {h} >_a \min_a(X_1',X_2',\cdots,X_{n-1}')\\
        &\implies  X_n' >_a \min_a(X_1',X_2',\cdots,X_{n-1}')\\
        &\implies \min_a(X_1',X_2',\cdots,X_n') = \min_a(X_1',X_2',\cdots,X_{n-1}') \\
        &\quad\quad \quad = X_2 >_a X_1
      \end{align*}

      Now, we apply the PR algorithm on \(X'\) under the valuation \(v_a\) to get a new allocation \(X''\). {We let \(b_1\) (resp. \(c_1\)) pick the most valued bundle from \(X''\) according to their valuation \(v_b\) (resp. \(v_c\)). If the bundles picked by \(b_1\) and \(c_1\) are different, then we have an EFX allocation. Otherwise, we rename that bundle as \(X''_n\).}
      From the property~2 of the PR algorithm, we also know that \(\min_a(X'') >_a \min_a(X') >_a X_1\). Therefore, \(\phi(X'') > v_a(X_1) = \phi(X)\). Thus we obtain a new almost EFX-feasible allocation with increased potential.

\medskip
    
   \textbf{Case 2: } The bundle \(X_n\setminus g_b\) is not the most favorite bundle of agent \(b_1\) \emph{and} bundle \(X_n\setminus g_c\) is not the most favorite bundle of agent \(c_1\). {Since,
    \(X_n\setminus g_b >_b \max_b(X_2,\cdots,X_{n-1})\) and 
    \(X_n\setminus g_c >_c \max_c(X_2,\cdots,X_{n-1})\) we have}
   
      $$X_n \setminus \{g_b\} <_b X_1 \cup \{g_b\}, \text{and }
      X_n \setminus \{g_c\} <_c X_1 \cup \{g_c\} $$

  In this case, we run the PR algorithm on \(\langle X_1\cup{g_b},\ X_n\setminus {g_b}\rangle\) under valuation \(v_b\) to get bundles \(Y_{n-1},Y_n\). Now the new allocation is \(X' = \langle X_1',X_2',\cdots,X_{n-1}',X_n'\rangle = \langle X_2,X_3,\cdots,Y_{n-1},Y_n \rangle\). 
  
  {We first show that bundles \(Y_{n-1}\) and \(Y_n\) are EFX-feasible for agent \(b_1\).}
  \begin{align*}
    \min_b(Y_{n-1},Y_n) &>_b \min_b((X_1\cup{g_b}), (X_n\setminus {g_b}))&&\\
            &= X_n \setminus \{g_b\} \quad \text{(\emph{Case 2} assumption)}&\\
            &>_b \max_b(X_2,\cdots,X_{n-1}) \ \text{(From~Equation~\ref{eq:1})}&\\
  \end{align*}

  \noindent Therefore, the bundles \(Y_{n-1}\) and \(Y_n\) are both EFX-feasible for agent \(b_1\).\\

  We let agent \(c_1\) choose their favorite bundle among \(Y_{n-1}\) and \(Y_n\), where \emph{w.l.o.g} let \(Y_n >_c Y_{n-1}\). {We now show that \(Y_n\) is EFX feasible for agent \(c_1\).} From the \emph{maximin} property of \(v_c\), we know the following:
  \begin{align*}
    Y_n &= \max_c(Y_{n-1},Y_n)& (\because Y_n >_c Y_{n-1})\\
        &\ge_c \min_c(X_1 \cup \{g_c\}, X_n \setminus \{g_c\})& \text{(\(v_c\) is MMS-feasible)}\\
        &= X_n\setminus \{g_c\}&\text{(\emph{Case 2} assumption)}\\
        &>_c \max_c(X_2,\cdots,X_{n-1})&\text{(From ~Equation~\ref{eq:2})}\\
  \end{align*}
  Therefore, the bundle \(Y_n\) is EFX-feasible for agent \(c_1\).\\
  Now, recall that the current allocation is \(X' = \langle X_2,X_3,\cdots,Y_{n-1},Y_n \rangle \). Depending on the envy from agent \(a_1\), we have the following three cases:

  \textbf{Case 2.1: } Agent \(a_1\) does not strongly envy \(Y_{n-1}\) and \(Y_n\). Since $X_2 <_a \cdots <_a X_{n-1}$, agents \(a_2,a_3,\cdots,a_{n-2}\) also does not strongly envy  \(Y_{n-1}\) and \(Y_n\). Thus, \(X'\) is an EFX allocation.
    
   \textbf{Case 2.2: } Agent \(a_1\) strongly envies both \(Y_{n-1}\) and \(Y_n\). Then,
 $$     Y_n >_a X_2,\quad
      Y_{n-1} >_a X_2, \quad
      X_3 >_aX_2 \text{(From ~Equation~\ref{eq:3})}$$
    Therefore, \(\min_a(X') = X_2 >_a X_1 = \phi(X)\). That is, the minimum has strictly increased. Now we run the PR algorithm on \(X'\) with  valuation \(v_a\) to get an almost EFX-feasible allocation \(X''\) with a potential value \(\phi(X'') > \phi(X)\).
    
   \textbf{Case 2.3: } Agent \(a_1\) strongly envies \(Y_{n-1}\) but not \(Y_n\). The other case is similar.\footnote{If agent \(a_1\) strongly envies \(Y_n\), then give \(Y_n\) to agent \(b_1\) and \(Y_{n-1}\) to agent \(c_1\). We know both \(Y_n\) and \(Y_{n-1}\) are EFX-feasible for agent \(b_1\). Thus we meet the invariant by making \(X_n''\) EFX-feasible for agent \(b_1\) instead of agent \(c_1\).}
  
    Let \(Y_{n-1}' \subseteq Y_{n-1} \) be such that \(Y_{n-1}'>_a X_2\) but \(Y_{n-1}'\setminus h <_a X_2\ \forall h\in Y_{n-1}'\). 
     Now consider the new allocation \(X'' = \langle X_1'',\cdots,X_{n-1}'', X_n'' \rangle = \langle X_2,\cdots, Y_{n-1}',\ Y_n\cup(Y_{n-1} \setminus Y_{n-1}') \rangle\).
Previously, \(Y_n\) was EFX-feasible for agent \(c_1\). Now, the value of this bundle has increased and values of other bundles have not increased. Therefore, the new bundle \(X_n''\) is EFX-feasible for agent \(c_1\).

    \sloppy The potential of the new allocation \(X''\) is \(\phi(X'') = \min_a(X_1'', X_2'', \cdots,Y_{n-1}') =  X_2 >_a  X_1 = \phi(X)\). That is, the potential value has increased. Now, if the bundles \(X_1'',X_2'',\cdots,X_{n-1}''\) are EFX-feasible for agents \(a_1,a_2,\cdots,a_{n-2}\), we are done.

    We know that bundles \(X_1'',\cdots,X_{n-2}''\) are EFX-feasible among themselves for agents \(a_1,a_2,\cdots,a_{n-2}\). By the construction of \(Y_{n-1}'\), it is clear that \(X_1'',X_2'',\cdots,X_{n-1}'' = Y_{n-1}'\) are EFX-feasible among themselves for agents \(a_1,a_2,\cdots,a_{n-2}\). Now, if \(X_1'',X_2'',\cdots,X_{n-1}''\) are EFX-feasible with respect to \(X_n''\), then all the invariant constraints are met and \(X''\) is a new almost EFX-feasible allocation with a higher potential value. Otherwise, if one of \(X_1'',X_2'',\cdots,X_{n-1}''\) is not EFX-feasible \emph{w.r.t.} \(X_n''\) according to valuation \(v_a\), then we have:
    \begin{align*}
      \exists h \in X_n''\text{ s.t. } X_n''\setminus {h} &>_a \min_a(X_1'',X_2'',\cdots,X_{n-1}'') = X_2\\
        &>_a \min_a(X_1,X_2,\cdots,X_n)\\
    \end{align*}
    That is, the overall minimum has increased. Now, we run the PR algorithm on \(X''\) with the valuation \(v_a\) to get a new allocation \(Z\). Let agent \(c_1\) pick their favorite bundle. From the property of the PR algorithm, we know that \(\phi(Z) >  \phi(X)\). Thus, we have a new almost EFX-feasible allocation with higher potential. This concludes the proof.
\end{proof}

%% file: sections/conclusion.tex
\section{Conclusion}
In this paper, we generalize the existing results in literature on EFX allocations to the setting when the number of distinct valuations is $k$, but the number of agents can be arbitrary. We give an EFX allocation with at most $k-2$ unallocated goods such that no agent envies the bundle of unallocated goods. We also show the existence of a complete EFX allocation under MMS-feasible valuations when all but two agents have identical valuations. The limitation of the technique used to prove Theorem~\ref{thm:kk2} is clear from \cite{efx_3}. At each step, our allocation Pareto dominates the previous allocations. As shown in \cite{efx_3}, even for three agents, there could be a partial allocation that Pareto dominates all complete allocations. So one cannot hope to reach a complete allocation using this technique. 
Reducing the number of unallocated goods or the number of outliers are challenging open questions.

%% file: sections/ack.tex
\section*{Acknowledgments}

We thank anonymous reviewers for their helpful comments. Vishwa Prakash HV acknowledges the support of TCS Research Scholar Fellowship. 